\documentclass[aps,twocolumn,nofootinbib]{revtex4-2}

\usepackage[english]{babel}
\usepackage[utf8]{inputenc}
\usepackage[normalem]{ulem}
\usepackage{amsmath,amssymb,amsfonts,physics,graphicx,xcolor,mathtools,dsfont,bm,hyperref}
\usepackage{tikz}
\usetikzlibrary{quantikz}
\usepackage{adjustbox}
\hypersetup{
    colorlinks=true,
    linkcolor={red!40!black},
    citecolor={blue!60!black},
    urlcolor={blue!50!black}
    }
\graphicspath{{fig_dos/}}
\definecolor{myred}{rgb}{0.85,0,0}
\definecolor{mygray}{rgb}{0.87, 0.87, 0.87}

\let\oldbibitem\bibitem 
\renewcommand{\bibitem}{
    \renewcommand{\doi}[1]{\texttt{\href{https://doi.org/##1}{doi:##1}}} 
    \let\bibitem\oldbibitem 
    \oldbibitem 
}

\newcommand{\exeter}{Department of Physics and Astronomy, University of Exeter, Stocker Road, Exeter EX4 4QL, United Kingdom}

\newcommand{\Deltak}{\hat{\Delta}_{k}}
\newcommand{\Sk}{\mathrm{S}_{k}}
\newcommand{\sansserif}[1]{{\fontfamily{cmss}\selectfont #1}}

\begin{document}

\title{Quantum topological data analysis via the estimation of the density of states}

\author{Stefano Scali}
\email{s.scali@exeter.ac.uk}
\affiliation{\exeter}
\author{Chukwudubem Umeano}
\affiliation{\exeter}
\author{Oleksandr Kyriienko}
\affiliation{\exeter}

\begin{abstract}
We develop a quantum topological data analysis (QTDA) protocol based on the estimation of the density of states (DOS) of the combinatorial Laplacian. Computing topological features of graphs and simplicial complexes is crucial for analyzing datasets and building explainable AI solutions. This task becomes computationally hard for simplicial complexes with over sixty vertices and high--degree topological features due to a combinatorial scaling. We propose to approach the task by embedding underlying hypergraphs as effective quantum Hamiltonians and evaluating their density of states from the time evolution. Specifically, we compose propagators as quantum circuits using the Cartan decomposition of effective Hamiltonians and sample overlaps of time--evolved states using multi--fidelity protocols. Next, we develop various post--processing routines and implement a Fourier--like transform to recover the rank (and kernel) of Hamiltonians. This enables us to estimate the Betti numbers, revealing the topological features of simplicial complexes. We test our protocol on noiseless and noisy quantum simulators and run examples on IBM quantum processors. We observe the resilience of the proposed QTDA approach to real--hardware noise even in the absence of error mitigation, showing the promise to near--term device implementations and highlighting the utility of global DOS--based estimators.
\end{abstract}

\maketitle


\section*{Introduction}
\label{sec:intro}

As the amount of data processed daily in the world exceeds 120 zettabytes~\cite{statista}, developing clever approaches to data analysis and storage plays an ever--growing role. There is particular interest in distilling and analyzing genuine features of datasets, such that the key information is captured and separated from irrelevant details. 

Topological data analysis (TDA) is a framework developed to extract topological information from datasets of varying dimension~\cite{Edelsbrunner_2002, Zomorodian_2004, CARLSSON_2005, Carlsson_2009, Wasserman_2018, Chazal_2021}, and thus learn features related to internal graph--type structures of the datasets. In TDA, we are given a set of points together with a distance metric. After choosing a filtration distance, a simplicial complex (a hypergraph of special type) is built on top of the data~\cite{Chazal_2021}. This complex is where the topological features of the data are encoded. However, when dealing with large datasets, imperfections in the form of missing data or noise are likely to appear. A strength of TDA is its resilience against these imperfections; this desirable feature has lead to an increasing number of applications for TDA in recent years~\cite{Wasserman_2018}. These applications range from the analysis of complex nonlinear fluid dynamics~\cite{Kram_r_2016} and nonlinear stochastic delay equations~\cite{Khasawneh_2016} to the classification of different pore geometries of materials~\cite{Lee_2017} and the imprints of cosmic structures~\cite{Pranav_2016}. 
One field where TDA--based techniques are proving particularly influential is the medical sector. Here, topological data analysis is used to represent brain activity maps~\cite{Saggar_2018} and to distinguish tumor from healthy regions in histology images by an interplay of persistent homology and convolutional neural networks ~\cite{Qaiser_2019}. TDA's influence extends to the field of machine learning, where it is used to enhance deep learning methods~\cite{Hensel_2021}. Further applications of TDA in machine learning are found for the classification of protein binding~\cite{Kovacev_Nikolic_2016}. 

Quantum algorithms aim to harness native correlations of quantum systems for achieving a scaling advantage over the classical counterparts~\cite{Bravyi_2018, Arute_2019}. 
In this direction, Lloyd et al. proposed a quantum algorithm to perform TDA (QTDA)~\cite{Lloyd2016} with a suggested exponential speed--up over the best classical algorithms. However, the proposed pipeline requires Grover's search~\cite{grover1996fast} to construct the simplices of the complex and quantum phase estimation (QPE)~\cite{kitaev1995quantum, Kitaev_1997} to estimate the kernel dimension of the combinatorial Laplacian, making it a costly algorithm that requires large--scale fault--tolerant quantum computers. 
Moreover, some of the resulting speed--ups were found to be quadratic under certain assumptions~\cite{gunn2019review, Berry2022}.
On the other hand, the estimation of (normalized) Betti numbers was shown to be \sansserif{DQC1}--hard, i.e. classically intractable~\cite{Gyurik_2022,cade2021complexity}. Furthermore, very recent work has demonstrated that problems involving clique complexes are $\mathsf{QMA}_1$--hard and contained in \sansserif{QMA}~\cite{king2023promise}, suggesting that the QTDA algorithm~\cite{Lloyd2016} is not \emph{dequantizable}~\cite{Tang_2019, gilyén2018quantuminspired, Chia_2020}. 
Crucially, these results leave open the question if instances of exponential quantum advantage can be found. 
\begin{figure*}
    \centering
    \includegraphics[width=.7\linewidth]{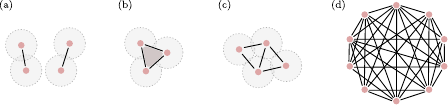}
    \caption{\textbf{Examples of simplicial complexes.} The dashed circles show the filtration scale $\epsilon$. (a) Simplicial complex with 4 vertices (0--simplices) and 2 edges (1--simplices). The complex has 2 connected components, thus $\beta_0=2$. In Appendix~\ref{app:cartan}, this complex is used to show an example of Cartan decomposition. (b) Simplicial complex with three 0--simplices (vertices), three 1--simplices (edges) and one 2--simplex (solid gray area). (c) Simplicial complex with $\beta_1=2$ used in the simulations and QPU runs. (d) Example of high dimensional ($N=10$) simplicial complex. Here, the filtration scale and the higher dimensional simplices are not shown for clarity. In Fig.~\ref{fig:fourier}, this complex is used as a proof--of--concept to show the effects of the frequency leaking phenomenon.}
    \label{fig:graphs}
\end{figure*}

In recent years, a particular interest has surrounded near--term applications, aiming to replace Grover's search and QPE with more hardware--friendly alternatives. Several alternative approaches have been proposed aiming to obtain better scalability and finding regimes of (exponential) quantum advantage~\cite{Ubaru2021, Akhalwaya2022, mcardle2022streamlined, Berry2022, Hayakawa_2022, schmidhuber2022complexitytheoretic} (see also the Discussion section in the end of the paper). At the same time, concerns about the existence of this advantage for non--trivial tasks of practical interest have also been raised~\cite{mcardle2022streamlined}.

In this landscape, we propose a protocol for QTDA that favors implementation on quantum hardware with limited resources, aiming to lower the barrier for its implementation. We do this thanks to the adoption of several techniques. First, we utilize an efficient subroutine for the time evolution of underlying the combinatorial Laplacian, such that the circuit complexity is reduced thanks to the use of optimal basis. Second, we use interferometric sampling protocols to estimate the density of states (DOS), moving part of the complexity towards the measurement stage. Lastly, we improve the Betti number estimation by applying classical signal processing methods, and retrieving these topological properties from analyzing evolution in the spectral (global) domain.
Rather than focusing exclusively on the kernel dimension of the combinatorial Laplacian, we show how different estimators help in reconstructing even its non--zero components (rank), thus showing the complementarity of our approaches to previous proposals.

In total, our QTDA protocol consists of the following main steps. We start by building a combinatorial Laplacian associated to a given simplicial complex. From this, we decompose the operator using the Cartan decomposition~\cite{Khaneja_2001}, resulting in a time--parametrized fixed--depth circuit. Other alternatives include Hamiltonian simulation methods, including Trotterization~\cite{Trotter_1959, Hatano_2005} and quantum signal processing~\cite{Low2017}. Next, we sample the evolution of the combinatorial Laplacian (being the effective Hamiltonian) at different times. We do this by means of interferometric protocols~\cite{Kyriienko_2020, Cortes_2022}, thus pushing the complexity towards the measurements while reducing circuit depth and register size. 
Lastly, we estimate the density of states~\cite{Schrodi2017} of the combinatorial Laplacian, to ultimately calculate its rank. This requires a post--processing of the trace evolution to recover the signal lost due to statistics (finite number of shots) and noise (real hardware). After the post--processing, we evaluate the Fourier components of the interpolated signal, calculate its rank, and estimate the Betti numbers. When tested on different simulation backends and real hardware (IBM quantum devices) we retrieve correct topological features. We stress that this has been achieve with no standard error mitigation techniques~\cite{Endo2018}, which could be implemented to further improve performance on hardware~\cite{Kim2023}.


\section{Protocol}
\label{sec:methods}
In what follows, we summarize the workflow to perform quantum topological data analysis via the density of states estimation. For the full derivation from a dataset to the Betti numbers, see Appendix~\ref{app:qtda}.
\begin{figure*}
    \centering
    \includegraphics[width=0.92\linewidth]{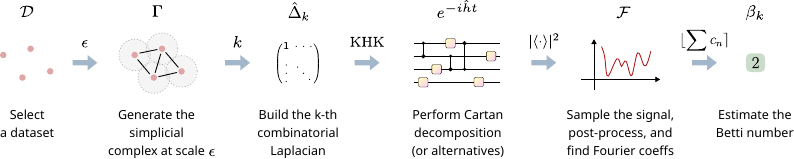}
    \caption{\textbf{Steps of the DOS--QTDA protocol.} A simplicial complex $\Gamma$ and the $k$--th combinatorial Laplacian $\Deltak$ are constructed from a dataset $\mathcal{D}$ at filtration distance $\epsilon$ and order $k$. In this work, we focus on the subsequent steps leading to the estimation of Betti numbers. These steps involve finding the Cartan decomposition $\hat{h}$ of the combinatorial Laplacian (using the KHK theorem), sampling the trace of the time evolution generated by $\hat{h}$, and post-processing the sampled signal to recover its Fourier components.}
    \label{fig:pipeline_dos}
\end{figure*}

Consider a dataset $\mathcal{D}$ consisting of $N$ points. Having chosen a filtration parameter $\epsilon$ and a metric $d$, we can build a simplicial complex $\Gamma$ and the associated $k$--th degree combinatorial Laplacian $\Deltak$. We show some examples of simplicial complexes in Fig.~\ref{fig:graphs}. The $k$-th Betti number of the dataset can be estimated as~\cite{Ubaru2021}
\begin{equation}\label{eq:betti_no}
    \beta_k = |\Sk| - \mathrm{rank}(\Deltak) ,
\end{equation}
where $|\Sk|$ is the number of $k$--simplices in the simplicial complex $\Gamma$. Knowledge of the simplicial complex $\Gamma$ and of the rank of $\Deltak$ are key to evaluating Betti numbers. The combinatorial Laplacian $\Deltak$ is obtained by the Hermitian boundary operator $\hat{B}$ written as a sum of fermionic operators. Then, using a Jordan--Wigner transformation on the operator, we recast it in terms of Pauli operators $\hat{X}$, $\hat{Y}$, and $\hat{Z}$ (see Appendix \ref{app:qtda}, following Ref.~\cite{Ubaru2021, Akhalwaya2022}). From here, the $k$--th combinatorial Laplacian $\Deltak$ is obtained as $\Deltak = \hat{P}_k \hat{\Delta}_\Gamma \hat{P}_k$, where $\hat{\Delta}_\Gamma$ is the combinatorial Laplacian obtained as the projection of the boundary operator $\hat{B}$ onto the $\Gamma$ simplicial complex via $\hat{P}_\Gamma$, and $\hat{P}_k$ is the projector onto the $k$--simplices (see details in Appendix~\ref{app:qtda}).

While we assume knowledge of the simplicial complex $\Gamma$, we propose to estimate the rank of $\Deltak$ via its density of states. To do this, we identify three main steps. First, we find the Cartan decomposition (or alternative Hamiltonian simulation approaches) of the combinatorial Laplacian. The Cartan decomposition approach has the advantage of a fixed circuit depth for different evolution times, convenient for the next steps. Second, we sample the unitary evolution at different times. We do this by using interferometric measurement routines. Lastly, we post process the sampled signal and evaluate its Fourier coefficients. The sum of the latter gives an estimate of $\mathrm{rank}(\Deltak)$. From here, we estimate the Betti numbers using Eq.~\eqref{eq:betti_no}. 
In Fig.~\ref{fig:pipeline_dos}, we show a schematic of the proposed DOS--QTDA protocol.

We proceed to present the detailed main steps of the proposed QTDA protocol based on the estimation of the density of states.


\subsection{Density of states}
\label{sec:dos}
The density of states (DOS) is of crucial interest when extracting the spectral information of an operator. This quantity, obtained as the Fourier transform of the unitary evolution of the operator, helps in all those tasks where the reconstruction of the microcanonical ensemble of the system is key. Notably, the power of the DOS estimation goes beyond the application of the Fourier transform, as different filters (in the form of integration windows/kernels/wavelets) can be used. Thus, the DOS performs well even in the presence of non--ideal measurements helping to extract more information by applying tailored estimators. Examples of this in practice include the accurate reconstruction of the DOS in 1D systems via tensor networks simulations~\cite{osborne2006renormalisationgroup, Schrodi2017}.

Consider a Hermitian operator $\Deltak$ which can be expressed in its eigenbasis as $\Deltak = \sum_{\nu} \lambda_\nu |\nu\rangle \langle \nu|$, where $\lambda_\nu$ are the eigenvalues and $|\nu\rangle$ the relative eigenvectors. The distribution of the eigenvalues of such an operator is expressed by its DOS $S(\omega)$,
\begin{equation}
    \label{eq:dos}
    S(\omega) = \sum_\nu \delta(\omega - \lambda_\nu) .
\end{equation}
Note that this quantity is also known as the eigenvalue density function~\cite{osborne2006renormalisationgroup}. From an appropriate variation of such a distribution, we estimate the rank dimension of the combinatorial Laplacian and consequently the Betti numbers via Eq.~\eqref{eq:betti_no}.

\subsubsection{Fourier components}
The density of states in Eq.~\eqref{eq:dos} is readily found by taking the Fourier transform ($\mathcal{F}$) of the trace of $\hat{U}(t)=e^{-i\Deltak t}$, the unitary time propagator generated by $\Deltak$. Thus, the main resource that allows the estimation of the DOS is the time evolution $\hat{U}(t)$ and this needs no controlled operations. We can express Eq.~\eqref{eq:dos} using a set of overlap measurements as
\begin{equation}
    \label{eq:dos_overlap}
    S_{k}(\omega) = \mathcal{F}\{ S(t) \} + r_{k} ,
\end{equation}
where $S(t) = \hat{U}(t)\hat{P}_{\Gamma\cap k} = \sum_{|\psi_j\rangle \in \Sk} \langle\psi_j| \hat{U}(t) |\psi_j\rangle = \tr_{|\psi_j\rangle}\{\hat{U}(t)\}$ with the states $|\psi_j\rangle$ selected randomly from the set $\Sk$ of $k$--simplices in the complex $\Gamma$ and $r_{k}=2^N-\mathrm{rank}(\hat{P}_{\Gamma \cap k})=2^N-\mathrm{rank}(\hat{P}_{\Gamma} \hat{P}_{k})$ is a constant estimated efficiently. Note that due to the time symmetric nature of the combinatorial Laplacian (see Appendix~\ref{app:symmetry}) the Fourier transform of the unitary evolution goes from a complex time domain to a real frequency domain. On top of this, since the combinatorial Laplacian is isospectral to the Hermitian positive semidefinite boundary operator $\hat{B}$, it has real non--negative eigenvalues (see Appendix~\ref{app:qtda}). By performing sampling (see Sec.~\ref{sec:sampling}) at different times on the set of states $|\psi_j\rangle \in \Sk$, we reconstruct $S(t)$ and, consequently, the DOS $S(\omega)$.
The spectrum of the combinatorial Laplacian, bounded from above (proven using the Gershgorin circle theorem~\cite{gunn2019review}), has frequencies in the range $0\leq\omega\leq N$. Thus, we can build a complete basis set for the trace evolution $S(t)$ by considering the first $N$ integer components. The coefficients relative to the single frequencies of the signal can be evaluated as
\begin{equation}\label{eq:fourier_coeffs}
    c_n = \frac{1}{T} \int_{-T/2}^{T/2} S(t) e^{-i 2\pi n t/T} \dd t ,
\end{equation}
where $T$ is the period of the function and $n\in \{1, 2, \cdots, N\}$ defines the set of discrete components of the spectrum. To avoid aliasing while covering the highest oscillating frequencies, we require the sampling frequency of the trace signal to be $f_\mathrm{s}=\lceil N\pi \rceil$, given by the Nyquist theorem. This limited sampling is possible thanks to the upper bound of the spectrum. By evaluating these $N$ coefficients and summing them up, we obtain an estimate of the rank of the combinatorial Laplacian $\Deltak$, that is, $\mathrm{rank}(\Deltak ) \simeq \sum_{n=1}^N c_n$. Betti numbers can be estimated using Eq.~\eqref{eq:betti_no} as
\begin{equation}\label{eq:betti_no_dos}
    \beta_k \simeq |\Sk| - \Big\lfloor\sum_{n=1}^N c_n\Big\rceil ,
\end{equation}
where $\lfloor \cdot \rceil$ indicates rounding to the closest integer. Note that the approximation is affected by several factors, namely the total number of sampling points (proportional to the total sampling period at a fixed sampling frequency), measurements statistics (number of shots), and noise (hardware--dependent). While the total number of sampling points and the measurement statistics are under our control in the form of increased resources required, we do not assume any noise control, that is, error mitigation. Thus, the signal $S(t)$ will be inevitably affected by noise. We propose a few simple steps of post--processing to recover the best possible signal with no introduction of data artifacts.

\subsubsection{Signal post--processing}
The post--processing routine consists of signal--boundary matching, mirroring, and interpolation.

\emph{Signal--boundary matching.}---The unitary evolution at $t=0$ resolves to the identity operator. Thus, we impose $\Re{S(0)}=\mathrm{dim}(\mathcal{H})=2^N$ and $\Im{S(0)}=0$. Shifting each sampling point by the difference between the corrected and the measured value at $t=0$, we restore the bias over the full signal. In the case of trigonometric interpolation, this matching is equivalent to requiring the Fourier coefficients to sum up to the Hilbert space dimension, $\sum_{n=0}^N c_n = 2^N$. Therefore, Eq.~\eqref{eq:betti_no_dos} can also be expressed as $\beta_k \simeq |\Sk| - (2^N - \lfloor c_0\rceil)$, where the zero--th coefficient $c_0$ represents the mean of the signal $S(t)$ in the sampled time interval, i.e. the kernel dimension.

\emph{Mirroring.}---The combinatorial Laplacian $\Deltak$ is time symmetric. Thus, we assume to have access to the back--propagated version of the time evolution imposing $\Re\big\{S(t^-)\big\} = \Re\big\{S(t^+)\big\}$ and $\Im\big\{S(t^-)\big\} = -\Im\big\{S(t^+)\big\}$, where $t^-$ and $t^+$ are the negative and positive sampling points respectively. This saves us half of the sampling cost.

\emph{Interpolation.}---We interpolate the post--processed signal $S(t)$ to obtain an integrable curve to use in Eq.~\eqref{eq:fourier_coeffs}. The interpolation effectively acts as a signal filter, the coefficients of the trigonometric interpolation coinciding with a ``clean'' discrete Fourier transform (DFT). Further details on the interpolation routine can be found in App.~\ref{app:interpolation}.

Therefore, after performing a limited amount of measurements, we reconstruct the trace signal $S(t)$, estimate the rank of the combinatorial Laplacian, and hence the Betti numbers (using Eq.~\eqref{eq:betti_no}).

\subsubsection{Frequency leaking}
\begin{figure}
    \centering
    \includegraphics[width=0.94\linewidth]{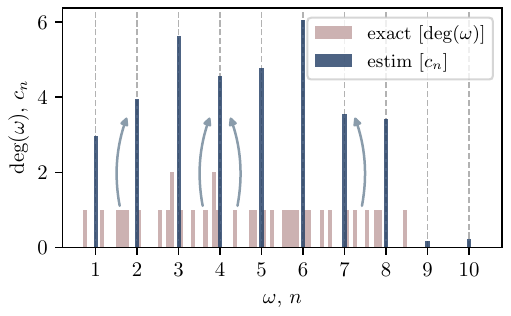}
    \caption{\textbf{Example Fourier components of the combinatorial Laplacian.} The simplicial complex studied ($N=10$) is shown in Fig.~\ref{fig:graphs}(d). The red bars refer to the exact spectrum of the combinatorial Laplacian, while the blue bars refer to the spectrum estimation with the Fourier coefficients. We analyze the spectrum of $\Delta_{2}$ and get the Betti number $\beta_2=9$. We obtain the result by estimating the rank of the combinatorial Laplacian as a the sum of the Fourier coefficients in Eq.~\eqref{eq:fourier_coeffs}. We subtract the rank (35.35) from the number of 2--simplices (44) in the complex and round to the closest integer (9). As seen in the figure, the leaking of the spectral components correctly captures the rank, thus the good estimation of the Betti number.}
    \label{fig:fourier}
\end{figure}

The combinatorial Laplacian is an integer matrix, i.e. its spectrum is composed of algebraic integer eigenvalues. For this reason, when dealing with systems of non--trivial dimensions, we find that non--integer frequencies \emph{leak} into some other integer component. This process is possible thanks to the completeness of the Fourier basis by which we construct the coefficients $c_n$. As a proof--of--concept, we show the effects of the leaking phenomenon in Fig.~\ref{fig:fourier}. We consider the simplicial complex $\Gamma$ shown in Fig.~\ref{fig:graphs}(d), study the unitary evolution generated by the combinatorial Laplacian $\Deltak$ and analyze the Fourier coefficients $c_n$. In this case, we use a B--spline interpolation to reconstruct $S(t)$~\cite{Knott_2000,julia_itp}.
The leaking phenomenon works reliably for spectra with a large spectral gap, however it suffers from those spectra whose spectral gap closes, thus approaching zero. In the latter case, the frequencies close to the zero might fall outside of the cover of the Fourier components and contribute to the kernel dimensionality. Combinatorial Laplacians with small spectral gap are a known regime of difficulty for quantum TDA~\cite{Ubaru2021,Berry2022,mcardle2022streamlined}.

We now show how to perform the time evolution generated by the combinatorial Laplacian as the effective Hamiltonian.


\subsection{Simulating dynamics: Cartan decomposition}
\label{sec:cartan}
The evaluation of the DOS relies on the simulation of the unitary evolution of the combinatorial Laplacian $\Deltak$. For this evolution, we can employ the Cartan decomposition of unitary operations initially proposed by Khaneja and Glaser~\cite{Khaneja_2001}, and we explicitly follow the algorithm of Kokcu et al.~\cite{Kokcu2022}. This method consists of three main steps that we summarize in the following. For a detailed discussion of the Cartan decomposition see Appendix~\ref{app:cartan}.

\emph{Basis.}---The first step is to find the basis for the Hamiltonian Lie algebra $\bm{\mathfrak{g}}$ of $\Deltak$, which is a subset of the full $\bm{\mathfrak{su}}(2^n)$ algebra. This can be done by searching for closure of the commutation relations of the Pauli terms composing the combinatorial Laplacian.

\emph{Decomposition.}---The second step is to find a Cartan decomposition of the Lie algebra $\bm{\mathfrak{g}}$. The involution associated with the time reversal symmetry of the combinatorial Laplacian ensures that, using the KHK theorem~\cite{Earp_2005}, we can always write
\begin{equation}
    \Deltak = \hat{K}\hat{h}\hat{K}^\dagger ,
\end{equation}
where $\hat{h}$ and the unitary operator $\hat{K}$ live in the orthogonal algebras defined by the Cartan decomposition.
Since our protocol relies on the evaluation of a trace, we can use its cyclic properties to obtain $S(t) = \tr\{\hat{U}(t)\} = \tr\{e^{-i\Deltak t}\} = \tr\{\hat{K}e^{-i\hat{h}t}\hat{K}^\dagger\}=\tr\{e^{-i\hat{h}t}\}$. Therefore, we can obtain $S(t)$ by sampling the time evolution generated by $\hat{h}$. Note that $\hat{h}$ is composed of commuting Pauli strings making its simulation easier. Importantly, the circuit depth to simulate $\hat{h}$ is independent of the simulation time and depends on the locality of the terms in $\hat{h}$.

\emph{Coefficients.}---The final step is to find the coefficients for the Pauli strings of $\hat{h}$. This is done via classical optimization. Depending on the simplicial complex this can be time consuming. In fact, there exists a direct correlation between the size of the Lie algebra $\bm{\mathfrak{g}}$ and the coefficients' optimization time~\cite{Kokcu2022}.
\begin{figure}
    \centering
    \includegraphics[width=0.75\linewidth]{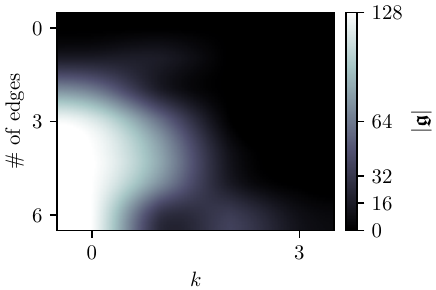}
    \caption{\textbf{Scaling of the Cartan Lie algebra.} The heatmap shows the interpolated average of the Lie algebra dimension of $\Deltak$ with respect to a set number of edges and homology group order $k$. The average is taken over the set of all simplicial complexes $\Gamma$ with $N=4$ vertices, the specified number of edges and homology group order $k$.}
    \label{fig:lie_algebra}
\end{figure}
In Fig.~\ref{fig:lie_algebra}, we show the dimension of the Lie algebra $|\bm{\mathfrak{g}}|$ with respect to the number of edges (1--simplices) of the simplicial complex studied and the homology group order $k$, for the case of $N=4$. The figure shows the average Lie algebra dimension over the set of possible simplicial complexes with a specific number of edges. It is clear from the figure that the combinatorial Laplacians of graphs which are near--complete generally have a larger Lie algebra dimension, hence a longer optimization time.

We know that the original $\Deltak$ lives in the computational basis reduced to $\Sk$. Subsequently, when calculating $S(t)$, we only need to measure overlaps involving the computational basis states in $\Sk$, as shown in Eq.~\eqref{eq:dos_overlap}. The states $\ket{\psi_j}\notin\Sk$ are orthogonal to the basis of $\Deltak$, therefore the contribution of these states to the overall trace can be accounted for without any actual measurement. This also applies to any approximate simulation of $\Deltak$ via Trotterization. However, this is no longer true when the Cartan decomposition is applied. Decomposing $\Deltak$ using the KHK theorem induces a scrambling of the basis. Even though the reduced Hamiltonian $\hat{h}$ contains the same spectral information as $\Deltak$, the states $\ket{\psi_j}\notin\Sk$ now possess a non--zero overlap with this new rotated basis of $\hat{h}$. Hence, Eq.~\eqref{eq:dos_overlap} does not apply when substituting $\Deltak$ for $\hat{h}$ and the measurements needed for trace estimation in Sec.~\ref{sec:sampling} require sampling over all $2^n$ computational basis states. This results in a disadvantageous sampling overhead particularly when $|\Sk|\ll2^n$.

As an example, at the end of Appendix~\ref{app:cartan}, we show the circuit of the Cartan decomposition relative to the simplicial complex in Fig.~\ref{fig:graphs}(a) and $k=0$.


\subsection{Sampling}
\label{sec:sampling}
After the state has undergone the Cartan evolution, we need to sample the trace of the evolution. We do this via interferometric measurements and hardware--friendly protocols, namely mirror measurements and the destructive SWAP test. We defer the details of the destructive SWAP test to the Appendix~\ref{app:swapd}.
\begin{figure*}
    \centering
    \includegraphics[width=0.84\linewidth]{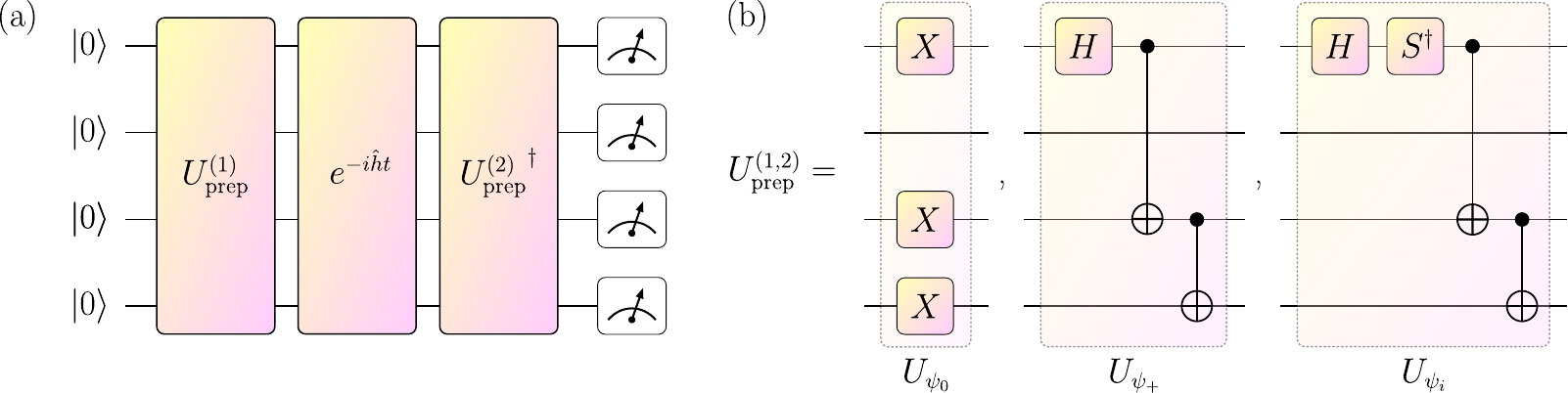}
    \caption{\textbf{Direct measurements and state preparation.} In panel (a), we show the routine of direct measurements to obtain the quantity $|\langle 0|(\hat{U}_\mathrm{prep}^{(2)})^\dagger e^{-i\hat{h}t} \hat{U}_\mathrm{prep}^{(1)}|0\rangle|^2$. In panel (b), an example set of circuits to prepare the states $\psi_0$, $\psi_+$, and $\psi_i$ respectively. The example represents the case of the hot state $\psi_0 = |1101\rangle$ and the entangled states $\psi_+ = |0000\rangle + |1101\rangle$ and $\psi_i = |0000\rangle + i|1101\rangle$.}
    \label{fig:direct_measurements}
\end{figure*}

Quantum protocols are generally accessed directly in the form of samples, resulting in histograms providing probabilities. The Hadamard test, an indirect approach giving access to expectation values, was introduced by Kitaev for quantum phase estimation~\cite{Kitaev_1997}. This approach adds a single ancillary qubit to fix a reference phase. In addition, it uses a set of controlled operations to control the target unitary. Because of the ancillary qubit and the large number of controlled operations, the implementation of the Hadamard test comes together with non--trivial noise. Since we aim to readily implement our protocol, we propose to use mirror measurements or the destructive SWAP test~\cite{Garcia_Escartin_2013, LaRose_2019, Suba__2019, Cincio_2018, Mitarai_2019}. These approaches dramatically reduce the number of controlled operations in the circuits, reducing the gap towards near--term applications.

To evaluate the real and imaginary parts of the expectations value of the trace operator, we use an interferometric measurement protocol~\cite{Kyriienko_2020, Cortes_2022, Bespalova2021}. While in an ancilla measurement protocol, the reference phase is encoded in an ancillary qubit, in the interferometric measurement protocol, the reference phase is identified with respect to some reference state. These protocols are sometimes referred to as single--fidelity and multi--fidelity, respectively. At the core, we construct different superposition states~\cite{Kyriienko_2020, Cortes_2022, Bespalova2021} of the reference state $|\psi_\mathrm{ref}\rangle$ and the target state $|\psi_0\rangle$. Then, we derive the expectation values of the combinatorial Laplacian via repeated measurements.

The real and imaginary parts of the expectation value $\langle \psi_0|e^{-i\hat{h}t}|\psi_0\rangle$ are obtained as,
\begin{align}
    \Re{\langle \psi_0|e^{-i\hat{h}t}|\psi_0\rangle} &= 2|\langle \psi_+|e^{-i\hat{h}t}|\psi_+\rangle|^2 \\
    &- \frac{1}{2}\left(1+|\langle \psi_0|e^{-i\hat{h}t}|\psi_0\rangle|^2\right) , \nonumber\\
    \Im{\langle \psi_0|e^{-i\hat{h}t}|\psi_0\rangle} &= -2|\langle \psi_i|e^{-i\hat{h}t}|\psi_+\rangle|^2 \\
    &+ \frac{1}{2}\left(1+|\langle \psi_0|e^{-i\hat{h}t}|\psi_0\rangle|^2\right) . \nonumber
\end{align}
In this way, the routine reduces to evaluating three overlap probabilities, namely $|\langle \psi_0|e^{-i\hat{h}t}|\psi_0\rangle|^2$, $|\langle \psi_+|e^{-i\hat{h}t}|\psi_+\rangle|^2$, and $|\langle \psi_i|e^{-i\hat{h}t}|\psi_+\rangle|^2$. Here, $|\psi_0\rangle$ is the ``hot'' state (target state) we want to evaluate the expectation value of, $|\psi_+\rangle = |\psi_\mathrm{ref}\rangle + |\psi_0\rangle$ is the superposition state between the reference state and the hot state and $|\psi_i\rangle = |\psi_\mathrm{ref}\rangle + i|\psi_0\rangle$ is the superposition state between the reference state and the dephased hot state. In our case, we can take the reference state $|\psi_\mathrm{ref}\rangle = |0\cdots0\rangle$, being the analogous of a fermionic vacuum state with no simplices associated and being orthogonal to the target hot states~\cite{Kyriienko_2020}. We refer to any of these preparations with the name of $\hat{U}_\mathrm{prep}^{(1,2)} = \hat{U}_{\psi_0}, \hat{U}_{\psi_+}, \hat{U}_{\psi_i}$ with the superscript $(1,2)$ indicating that the two state--preparation unitaries in the routine can be different (specifically, in the case of $|\langle \psi_i|e^{-i\hat{h}t}|\psi_+\rangle|^2$).

While the unitary $\hat{U}_{\psi_0}$ prepares the hot pure states of the computational basis (sampling the $2^N-1$ hot states), the unitary $\hat{U}_{\psi_+}$ and $\hat{U}_{\psi_i}$ prepare the relative GHZ--like state entangled with the vacuum state. To implement $\hat{U}_{\psi_0}$ a single string of $X$ gates is sufficient. To implement the entangled states, a register is initialized with a Hadamard gate and a cascade of CNOTs follows to build the hot components. A state of the form $|0\cdots0\rangle + |m\mathrm{-hot}\rangle$ is created, where the $m$-hot component draws from the hot states. An example preparation for the states $\hat{U}_{\psi_0}|0000\rangle = |1101\rangle$, $\hat{U}_{\psi_+}|0000\rangle = |0000\rangle + |1101\rangle$, and $\hat{U}_{\psi_i}|0000\rangle = |0000\rangle + i|1101\rangle$ are shown in Fig.~\ref{fig:direct_measurements} (b).

The mirror measurement is the most fundamental measurement protocol being the easiest to implement (often used in quantum kernel measurements~\cite{Paine2023}). As shown in Fig.~\ref{fig:direct_measurements}, to evaluate $|\langle 0|(\hat{U}_\mathrm{prep}^{(2)})^\dagger e^{-i\hat{h}t} \hat{U}_\mathrm{prep}^{(1)}|0\rangle|^2$, we prepare the state with $\hat{U}_\mathrm{prep}^{(1)}$, run the time evolution $e^{-i\hat{h}t}$, and unprepare the state with $(\hat{U}_\mathrm{prep}^{(2)})^\dagger$. The return probability corresponds to the quantity sought. In Fig.~\ref{fig:direct_measurements}(a) we show an example circuit for the mirror measurement routine.
\begin{figure*}
    \centering
    \includegraphics[width=\linewidth]{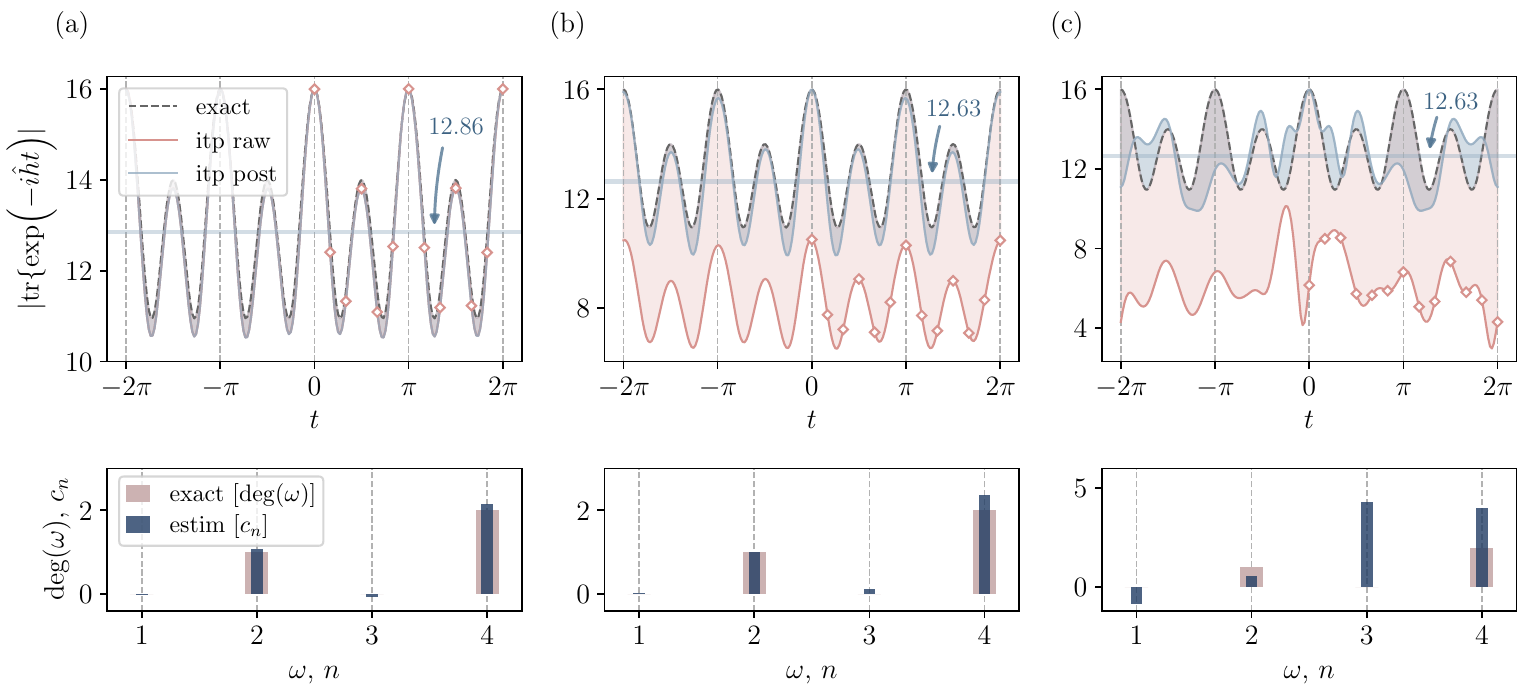}
    \caption{\textbf{Noiseless simulation, noisy simulation, and QPU run.} Both simulations and the hardware run refer to the simplicial complex in Fig.~\ref{fig:graphs}(c) and $k=1$. In the top row, the dashed black curves refer to the exact time evolution of the combinatorial Laplacian, the solid red curves refer to the interpolated raw sampled signals (sampled points indicated by the red diamonds), while the solid blue curves are the relative post-processed interpolations. The colored areas indicate the signal difference between exact/raw signals (red area) or exact/post-processed signals (blue area). The horizontal blue line corresponds to the estimated $c_0$. In the bottom row, the large red bars refer to the exact spectrum of the combinatorial Laplacian, while the narrow blue bars refer to the spectrum estimation with the Fourier coefficients.
    In panel (a), the noiseless simulation produces an excellent output signal that barely benefits from the post-processing routine. The spectrum and rank of the combinatorial Laplacian $\hat{\Delta}_1$ are correctly estimated. In (b), we show the noisy simulation (noise model \sansserif{FakeWashingtonV2} on \sansserif{Qiskit}~\cite{Qiskit}) results in a strong but uniform signal loss. The post-processing routine here recovers the vast majority of the signal resulting in a good spectrum estimation, thus rank of the combinatorial Laplacian and Betti number. In (c), the hardware run (on IBM Brisbane QPU) introduces even stronger signal loss, this time accompanied by decoherence. While the post--processing recovers the noise bias, there is an inevitable scrambling of the spectrum information. On the other hand, the kernel information remains reliable enough to correctly estimate the Betti number.}
    \label{fig:runs}
\end{figure*}

Note that both mirror measurements and the destructive SWAP test push the readout complexity towards the sampling process requiring $1/\varepsilon^2$ calls to the quantum computer to achieve an estimation with a precision of order $\varepsilon$~\cite{Wang_2019, Mitarai_2019}.


\section{Results}
\label{sec:results}
In this section, we focus on the implementation of the routine just presented, included noiseless simulations, noisy simulations, and real hardware runs. All simulations and real hardware runs are implemented using the \sansserif{Python} package \sansserif{Qiskit}~\cite{Qiskit} and IBM Quantum platform for flexibility and hardware compatibility. The rest of the pipeline is built on \sansserif{Julia} language~\cite{bezanson2017julia} for an improved workflow and performance, including data point analysis, simplicial complex creation, signal post--processing, and Betti number estimation via DOS. To find the Cartan decomposition parameters, we call the \sansserif{Python} package \sansserif{cartan--quantum--synthesizer}~\cite{cartan}. We use different \sansserif{pytket}~\cite{Sivarajah_2020} passes onto the time--parametrized (using static UUIDs) version of the Cartan circuit to optimize it while maintaining the same number of gates. We built ad--hoc parsing methods to interface the two branches (\sansserif{Python}, \sansserif{Julia}) of the pipeline.

For our runs, we consider the simplicial complex in Fig.~\ref{fig:graphs}(c). This complex has four vertices and five 1--simplices with no higher dimensional simplices. We study the Betti number corresponding to $k=1$ that is, $\beta_1=2$. The eigenvalues of the combinatorial Laplacian $\hat{\Delta}_1$ are $\lambda_1=0, \lambda_2=0, \lambda_3=2, \lambda_4=4, \lambda_5=4$, thus $\mathrm{rank}(\hat{\Delta}_1)=3$ and $\mathrm{dim}(\mathrm{ker}(\hat{\Delta}_1))=2$.

To begin with, we find the Cartan decomposition associated with the combinatorial Laplacian $\hat{\Delta}_{1}$. The dimension of the Lie algebra $\mathfrak{g}$ is $\mathrm{dim}(\mathfrak{g}) = 128$. The full gate decomposition includes $113$ gates (of which $50$ are CNOTs), optimized via \sansserif{pytket}~\cite{Sivarajah_2020} to $49$ gates ($16$ CNOTs). Once the decomposition is found, we run the circuit at different times and sample in the range $[0, 2\pi]$. In this case, it is sufficient to have $f_s=13$ to cover for the highest frequency. We run $1000$ shots per overlap measurement, sufficient to resolve the spectrum degeneracy for this dimensionality in the simulations. We run $10,000$ shots on the hardware run to limit further the statistical variance in the case of noisy QPUs. We sample the signal on all $2^4$ canonical basis states due to the Cartan scrambling of the basis, ensuring no information is lost. Finally, the multi--fidelity sampling requires us to run the protocols three times for each expectation value $\langle \psi_j|e^{-i\hat{h}t}|\psi_j\rangle$.

\emph{Noiseless simulation.}---In Fig.~\ref{fig:runs}(a), we show the noiseless run of this protocol. It is clear that the raw sampled signal is already good enough to correctly approximate the spectrum of the combinatorial Laplacian. No notable information recovery is obtained by the post--processing routine as the \emph{itp raw} and the \emph{itp post} curves fall on top of each other. In this case, noise is not a factor and we run the destructive SWAP test as shown in Appendix~\ref{app:swapd}. We obtain a nearly perfect spectrum (where slight deviations are caused by a limited statistical sampling) with $\mathrm{rank}(\hat{\Delta}_1)=3.16$. The Betti number is therefore correctly estimated to be $\beta_1=2$.

\emph{Noisy simulation.}---In Fig.~\ref{fig:runs}(b), we show the noisy simulation obtained with the \sansserif{FakeWashingtonV2} noise model. This model is built from the corresponding Washington IBM quantum processing unit (QPU) with the 127--qubit Eagle architecture~\cite{ibm_resource}. For this simulation and the following real--hardware run, we choose the mirror measurement scheme as the most efficient in the presence of noise. In this case, the raw sampled signal suffers from a substantial signal loss at all times while it does not decohere significantly. After the post--processing, we recover most of the signal as shown by the red area between the \emph{itp raw} and the \emph{itp post} curves. The sum of the reconstructed Fourier coefficients $c_n$ gives $\mathrm{rank}(\hat{\Delta}_1)=3.49$. Again, the Betti number is estimated to be $\beta_1=2$.

\emph{QPU run.}---Now, we show the robustness of the proposed protocols by running the routine on a real hardware with limited connectivity. In Fig.~\ref{fig:runs}(c), we show the run obtained on the Brisbane IBM QPU based on the 127--qubit Eagle architecture. The post--processed signal decoheres significantly with respect to the exact signal causing the spectrum to deviate from the exact result. In addition, while the combinatorial Laplacian $\Deltak$ has real, non--negative algebraic integer eigenvalues by construction (see Appendix~\ref{app:qtda}), negative eigenvalues (Fourier coefficients) are present in the reconstructed spectrum. The sum of the first 4 coefficients in this case is $\mathrm{rank}(\hat{\Delta}_{1}) = 8.0$, affecting the correct estimation of the Betti number. Despite all these signal artifacts, the reconstructed zero coefficient $c_0$ (accounting for the signal bias, i.e. the kernel dimensionality) does not undergo major changes. As a consequence, the rank of the combinatorial Laplacian can be well estimated using $\mathrm{rank}(\Deltak) \simeq 2^N - \lfloor c_0\rceil$ and with it, the Betti number. For this run, we obtain $c_0=12.63$ resulting in $\mathrm{rank}(\hat{\Delta}_{1}) = 3.37$. The Betti number is therefore estimated to be $\beta_1=2$.

Importantly, the zero--coefficient serves as a reliable method to estimate the Betti numbers in the case of a signal strongly affected by noise. Given the signal--boundary conditions imposed in Sec.~\ref{sec:dos}, all coefficients sum up to the Hilbert space dimension. Therefore, the frequencies leaking into the higher--end of the spectrum and not summed in the rank evaluation $\mathrm{rank}(\Deltak) \simeq \sum_{n=1}^N c_n$ are still picked--up by the zero coefficient as $\mathrm{rank}(\Deltak) \simeq 2^N - c_0$. Alternatively, a re--sampling of the signal (up--sampling followed by down--sampling) can be performed to make the highest frequencies alias into the frequencies within $n=\{1, \ldots, N\}$.
 
All runs are set to \sansserif{optimization=0} in \sansserif{Qiskit}~\cite{Qiskit} to avoid circuit depth overwriting of the Cartan circuit. While with this choice we pay a little price in terms of circuit optimization, this helps us to maintain a homogeneous bias throughout the sampling of the different basis states at different times. At the same time, it helps reducing the need for any sort of Loschmidt echo or zero noise extrapolation~\cite{Temme_2017, Li_2017, majumdar2023best}, limiting the post--processing to the steps described in Sec.~\ref{sec:dos}.

As already mentioned, these results aim to show the robustness of the protocol in different scenarios. An immediate performance boost and improved accuracy can be found by running the routine on fully--connected platforms or applying error mitigation routines. Further improvements can be achieved by using tailored filters or interpolating techniques (see Appendix~\ref{app:interpolation}).

Another way to improve our results is to sample the evolution up to longer times. This would allow us to capture slower frequencies. Thus, one can better resolve the Fourier coefficients for periods multiple of $2\pi$ and later remap the frequencies within the spectrum domain. As shown in Appendix~\ref{app:interpolation}, the period depends on the single eigenvalues. The multiple of periods needed to include the spectral gap frequency $\lambda_\mathrm{gap}$ is $\lceil \lambda_\mathrm{gap}^{-1} \rceil$. While this is not enough for a perfect spectrum reconstruction, it is a sensible choice to allow the leaking procedure to bound all the frequencies of the spectrum. Note that this process of longer time evolution (and relative sampling) naturally leads to the definition of the continuous Fourier transform in the infinite--time limit while maintaining degeneracy information.

We finally note that increasing the sampling frequency in the single period does not usually result in a refined output signal. To increase the accuracy of the readout, it is suggested to refine the statistics of the single time points that is, to increase the number of shots per sampling point.


\section{Discussion}
\label{sec:discussion}
Topological data analysis has demonstrated a significant potential when translated into a quantum processing workflow, and much work has been produced to date. Here, we briefly discuss recent quantum TDA approaches for extracting Betti numbers.

The $k$-th Betti number $\beta_k$ can be expressed as in Eq.~\eqref{eq:betti_no}, thus as the kernel cardinality of $\Deltak $. In Ref.~\cite{Berry2022} the authors suggest to use quantum phase estimation (QPE) for an ensemble of states in $\hat{P}_{\Gamma\cap k}$ for the boundary operator $\hat{B}_\Gamma$, isospectral to $\Deltak$. The authors evaluate the frequency of getting zero eigenvalues, hence the kernel cardinality, by repeated QPE. This is an efficient approach in term of scaling and T gate counts. However, it requires the implementation of a controlled $\exp\big(-i\hat{B}_\Gamma t\big)$ and various subroutines that are suited to large scale fault--tolerant quantum computers, and being less suitable for near--term and mid--term devices with limited resources (including early fault--tolerant ones).

A different approach introduced in Refs.~\cite{Ubaru2021,Akhalwaya2022} relies on stochastic rank estimation using the identity $\mathrm{rank}(\Deltak ) \equiv \tr \{ \Theta(\Deltak ) \}$, where $\Theta(x)$ is the Heaviside function. The authors then use the Chebyshev decomposition of the estimator, and evaluate the Chebyshev moments for the different states in a simplex, $\langle s_\ell| T_j (\Deltak ) |s_\ell \rangle$, for all $\ell, k, j$. This is performed via probabilistic projections with $N$ ($N^2$) ancillary qubits with (without) mid--circuit measurements. Here, the depth of the QTDA protocol is kept shallow at the expense of a largely increased number of samples required to estimate the moments. This can be the bottleneck of the approach, especially for hardware architectures with slower sampling rates.

Positioning our work, the DOS--based QTDA protocol tries to balance the circuit depths while still leveraging time evolution and keeping the measurement overheads smaller. This allows hardware tests and can be brought forward into the early fault--tolerant or ISQ era~\cite{ISQ}.

As a final note, we believe that there is scope in the field to advance quantum protocols for topological data analysis, as well as reducing their cost, facilitating hardware implementations. Some of these advances may include novel representations of the combinatorial Laplacian~\cite{mcardle2022streamlined} which could simplify the time dynamics, methods utilizing the symmetries of the simplicial complex, or perhaps new procedures which bypass the evolution all together. Finally, a complete landscape of the quantum suitability of simplicial complexes, together with their applications, remains desirable. These are some of the open challenges that could unlock the full potential of QTDA in enhancing data analysis and explainable AI.

\section{Conclusions}
\label{sec:conclusions}

In this work we have proposed the quantum protocol for topological data analysis based on the density of states estimation. While prior work often targeted fault tolerant quantum computing settings with no end--to--end simulation, here we focused on algorithms that can be run on available hardware. Our protocol pipeline involves three main steps. First, we use the time evolution generated by the combinatorial Laplacian, compiling it with the help of the Cartan decomposition. Second, we sample the overlaps of the evolved and reference states using a multi--fidelity approach. We concentrate on overlap estimators that minimize the number of qubits and that are less prone to noise, noting that other options are available depending on the hardware noise properties. Lastly, we estimate the density of states of the post--processed sampled signal, evaluating a set of Fourier coefficients and estimating the Betti numbers.

We showcased examples of DOS--QTDA with simulation and QPU runs. We run our protocol on noiseless/noisy simulators and the IBM Eagle architecture (Brisbane QPU). The compact circuit given by the Cartan decomposition, the near--term routines, and the post processing worked together to give good results in the chosen dimensionality. The resiliency of the full protocol presented here is supported by the use of a platform with limited connectivity and in the absence of error mitigation protocols.


\begin{acknowledgments}
We would like to thank Ravindra Mutyamsetty, Zhihao Lan and Okello Ketley for useful discussions. The authors acknowledge the support from the Innovate UK ISCF Feasibility Study project number 10030953 granted under the ``Commercialising Quantum Technologies'' competition round 3.

The views expressed are those of the authors and do not reflect the official policy or position of IBM and the IBM Quantum team.
\end{acknowledgments}


\section{Appendix}
\label{sec:appendix}

\subsection{Topological data analysis and quantum embedding}
\label{app:qtda}
The topological features of a dataset are hidden in the associated simplicial complex at a certain filtration distance. By studying the kernel of the combinatorial Laplacian associated with the simplicial complex, one can learn the Betti numbers of the dataset topology. Following Ref.~\cite{Ubaru2021, Akhalwaya2022}, we summarize the main steps to obtain the combinatorial Laplacian. We direct the reader to the original works just referenced for more details on the construction of the single elements.

We start with a dataset $\mathcal{D}$, collection of $N$ points (vertices) with $M$ coordinates, $\mathcal{D} = \{(x^{(1)}_i, \ldots, x^{(M)}_i)\}_{i=1}^{N}$. Here, $x^{(j)}_i$ is the $j$-th coordinate relative to the $i$-th point. Given a scale (filtration) parameter $\epsilon$ and a distance metric $d$, we can build a simplicial complex $\Gamma$ by collecting simplices of vertices that fall within the distance $d\leq\epsilon$. Depending on the distance metric chosen, different definitions of complexes can be made, notable examples of which are the Vietoris--Rips or C\v{e}ch complexes. 
Once a simplicial complex $\Gamma$ is specified, we build the associated combinatorial Laplacians and study their ranks, thus obtaining information on the topology of the dataset. This represents a formidable challenge for complexes with large $N$. Since the number of $k$-simplices $|\Sk|$ grows combinatorially with $N$, performing a full filtration is a challenging task. Given this scaling, one would naturally look at quantum systems as a possible solution. To encode combinatorial Laplacians as quantum operators, we use a standard approach discussed in Refs.~\cite{Lloyd2016,Berry2022,Ubaru2021,Akhalwaya2022}. The procedure relies on associating the vertices of the complex (0--simplices) with the one--hot computational basis states, $|100\cdots0\rangle$, $|010\cdots0\rangle$, $|001\cdots0\rangle$ etc. The $k=1$-simplices are encoded with the two--hot computational basis states, $|110\cdots0\rangle$, $|101\cdots0\rangle$, $|011\cdots0\rangle$ and so on. The simplices of all dimensions can be encoded that is, up to $k=N-1$. To build the combinatorial Laplacian, we need to form the restricted boundary operator $\hat{\partial}_k$ as
\begin{equation}
    \hat{\partial}_k|s_k\rangle = \sum_{\ell = 0}^{k-1} (-1)^\ell |s_{k-1}(\ell)\rangle .
\end{equation}
This linear operator connects the states relative to the $k$-simplices to the states relative to the $(k-1)$-simplices, producing an oriented boundary map. When all boundary operators over different $k$ are summed, we obtain the full boundary map $\hat{\partial} = \oplus_k \hat{\partial}_k$, which can be written as a sum of fermionic operators, $\hat{a}_i$ s.t. $\{\hat{a}_i, \hat{a}_j^\dagger \} = \delta_{i,j}$ for any $i,j$. Using the Jordan--Wigner transformation, we can recast the Hermitian boundary operator $\hat{\partial} + \hat{\partial}^\dagger$ as
\begin{align}\label{eq:Bop}
    \hat{B} :&= \hat{\partial} + \hat{\partial}^\dagger = \sum_{i=1}^N (\hat{a}_i + \hat{a}_i^\dagger) \\ \nonumber
    &= \hat{X}_1 + \hat{Z}_1 \hat{X}_2 + \hat{Z}_1 \hat{Z}_2 \hat{X}_3 + \cdots + \left(\prod_{i<N} \hat{Z}_i \right) \hat{X}_N,
\end{align}
where $\hat{X}_i$, $\hat{Y}_i$, $\hat{Z}_i$ are Pauli operators at sites $i=1, 2, \cdots, N$. Note that $\hat{B}^2 = \mathbb{I}$, being an involutory operator. The total boundary operator is unique for all complexes and simplices, however, its form changes depending on the basis in which the problem is encoded. Note that, the definition of the Hermitian boundary operator defined as $\hat{B} := \hat{\partial} + \hat{\partial}^\dagger$ makes it a positive semidefinite operator, thus having real non--negative eigenvalues. The combinatorial Laplacian, isospectral to $\hat{B}$, is then formed from the products of the projected boundary operators, which reads~\cite{Ubaru2021,Akhalwaya2022}
\begin{equation}
    \hat{\Delta}_{\Gamma} := \hat{P}_{\Gamma} \hat{B} \hat{P}_{\Gamma} \hat{B} \hat{P}_{\Gamma},
\end{equation}
where $\hat{P}_{\Gamma}$ is a projector on all $k$-simplices in the complex $\Gamma$. For instance, the complex in Fig.~\ref{fig:graphs}(a) is specified as $\hat{P}_{\Gamma} = \sum_{i=1}^4 \hat{P}_i^{(k=0)} + |1100\rangle \langle 1100| + |0011\rangle \langle 0011|$ where $\sum_{i=1}^N \hat{P}_i^{(k=0)}$ is the projector on all the one--hot computational basis representing the vertices. As a further example, the complex in Fig.~\ref{fig:graphs}(b) is $\hat{P}_{\Gamma} = \sum_{i=1}^3 \hat{P}_i^{(k=0)} + |110\rangle \langle 110| + |101\rangle \langle 101| + |011\rangle \langle 011| + |111\rangle \langle 111|$. Finally, as we want to learn the topological properties for a specific $k$, we need to project the Laplacian onto the states relative to the $k$-simplices,
\begin{equation}
\label{eq:DeltaGk}
    \Deltak  := \hat{P}_{k} \hat{P}_{\Gamma} \hat{B} \hat{P}_{\Gamma} \hat{B} \hat{P}_{\Gamma} \hat{P}_{k}.
\end{equation}
Since the projectors $\hat{P}_{\Gamma}$ and $\hat{P}_{k}$ commute, $[\hat{P}_{\Gamma}, \hat{P}_{k}] = 0$, being formed in the same basis, we can combine them in the projection of the intersection defined as $\hat{P}_{\Gamma \cap k} := \hat{P}_{\Gamma} \hat{P}_{k}$. The combinatorial Laplacian \eqref{eq:DeltaGk} can be seen as the unitarily transformed (by $\hat{B}$) complex projector, and projected onto $\hat{P}_{\Gamma \cap k}$. Having an expression for the combinatorial Laplacian $\Deltak$, we can estimating the Betti numbers by using
\begin{equation}
    \beta_k = |\Sk| - \mathrm{rank}(\Deltak ) .
\end{equation}


\subsection{Interpolation of the signal}
\label{app:interpolation}
To obtain kernel information via the analysis of the DOS, we resort to the evaluation of the Fourier components extracted from the signal $S(t)$. Once we sampled the signal, we perform the post--processing routine shown in Sec.~\ref{sec:dos}. Here, we focus on the last step of such routine, the interpolation of the signal $S(t)$.

Given the unitary nature of the problem, we choose to interpolate with the trigonometric polynomial,
\begin{equation}\label{eq:itp_trig}
    S(t) = \sum_{m=-(M-1)/2}^{(M-1)/2} \frac{X_m}{M} e^{i2\pi m t} ,
\end{equation}
where the $X_i$ are the DFT components of the shifted DFT vector (in order: negative components, zero component, positive components) of the sampled signal with $M=2f_s-1$ to satisfy the periodicity of the DFT. This interpolation is then followed by the evaluation of the Fourier coefficients using Eq.~\eqref{eq:fourier_coeffs}. Importantly, we truncate the signal sampling after a time $t=2\pi$, in most cases reducing the information acquired of the slowest frequencies. The period of the trace evolution is, in fact, equal to the least common multiple of the single periods, that is, $T_{S(t)}=\mathrm{LCM}(2\pi\lambda_\mathrm{gap}^{-1}, \, 2\pi\lambda_2^{-1},\, \cdots,\, 2\pi\lambda_{\Bar{M}}^{-1})$ where the unique $\Bar{M}$ non--zero eigenvalues $\lambda_i$ are in increasing order. Here, $\lambda_\mathrm{gap}$ is the smallest non--zero eigenvalue, i.e. the spectral gap frequency. A signal sampled up to a time shorter than $T$ will not match the periodicity at time $t=2\pi$ and needs to be adjusted.

In the case of unmatched signal periodicity, different interpolation techniques can be applied to obtain ad--hoc results. For example, the piecewise nature of B--splines comes in help when having a signal with closing spectral gap, i.e. where a long sampling time is needed. In this case, B--splines modify the signal only in the last quadrant of the interpolation by matching the last sampling point to the first [together with its derivative(s)]. On the other hand, the signal sampled at the previous time points remains unchanged, i.e. the closest to the original readout. Different interpolations effectively acts as different signal filters; the trigonometric interpolation coinciding with a ``clean'' discrete Fourier transform (DFT). The latter finds its greatest utility when dealing with periodic functions, therefore with those functions with reasonably large (in the sense of time sampling) spectral gap. We leave the study of alternative interpolating and filtering techniques~\cite{Harris_1978, Nuttall_1981, Heinzel2002SpectrumAS} to future work.


\subsection{Proof of the time reversal symmetry of the combinatorial Laplacian}
\label{app:symmetry}
As shown in Ref.~\cite{Kokcu2022}, Hamiltonians that possess the time reversal symmetry consist of Pauli strings with an even number of $\hat{Y}$ matrices, or alternatively the Hamiltonians are real, symmetric matrices.
The general form for $\Deltak $ is given in Eq.~\eqref{eq:DeltaGk}. The breakdown of $\hat{B}$ in terms of Pauli operators is given in Eq.~\eqref{eq:Bop}, and we know that $\hat{P}_\Gamma$ is a diagonal projector onto the computational basis elements, so for any $n$-qubit complex it can always be written as
\begin{equation}\label{eq:projs}
    \hat{P}_\Gamma = \alpha\mathds{1}+\sum_i\beta_i\hat{Z}_i+\sum_{i<j}\gamma_i\hat{Z}_i\hat{Z}_j+\cdots+\omega \hat{Z}_1 \cdots \hat{Z}_n ,
\end{equation}
where $\{\alpha_i,\beta_i, \ldots, \omega\}$ are real coefficients. 

\begin{figure*}
    \centering
    \includegraphics[width=0.86\linewidth]{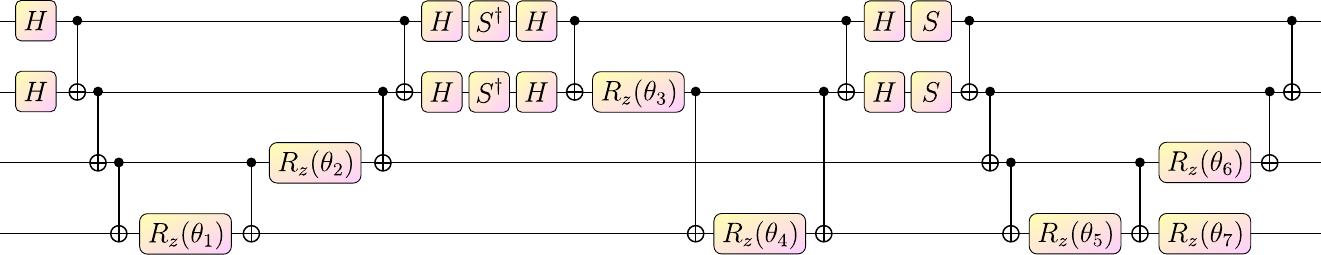}
    \caption{\textbf{Example of Cartan decomposition circuit.} The circuit represents the unitary evolution $e^{-i\hat{h}t}$. The Hamiltonian $\hat{h}$ is found from the Cartan decomposition of $\Deltak $, where $\Gamma$ is the complex in Fig. \ref{fig:graphs}(a). Note that we omitted a global phase $e^{i\theta_8}$ since it does not affect the spectrum.}
    \label{fig:cart_circ}
\end{figure*}

Hence, neither $\hat{B}$ or $\hat{P_\Gamma}$ contain any strings containing $\hat{Y}$ terms, so they are both symmetric operators. By definition, a matrix $\hat{A}$ is symmetric if $\hat{A}^T=\hat{A}$. It is also important to note that for two symmetric matrices $\hat{A}$ and $\hat{B}$,
\begin{equation}
    (\hat{A}\hat{B}\hat{A})^T = (\hat{A}(\hat{B}\hat{A}))^T = (\hat{B}\hat{A})^T\hat{A}^T = \hat{A}^T\hat{B}^T\hat{A}^T = \hat{A}\hat{B}\hat{A} .
\end{equation}
From this, we can conclude that the product $\hat{B}\hat{P}_\Gamma\hat{B}$ is also symmetric.

We can proceed using similar arguments. The combined projector $\hat{P}_{\Gamma \cap k}=\hat{P_k}\hat{P_\Gamma}=\hat{P_\Gamma}\hat{P_k}$ is again a diagonal projector with the form given in Eq.~\eqref{eq:projs}. Hence $\hat{P}_{\Gamma \cap k}$ is symmetric, and since $\hat{B}\hat{P}_\Gamma\hat{B}$ is symmetric, $\hat{P}_{\Gamma \cap k}\hat{B}\hat{P}_\Gamma\hat{B}\hat{P}_{\Gamma \cap k}=\Deltak $ is also symmetric.

Thus, for any possible complex and simplex order, the corresponding combinatorial Laplacian is a symmetric matrix. We know that, $\hat{B}$, $\hat{P_\Gamma}$ and $\hat{P_k}$ are all Hermitian matrices, so by the same arguments we can show that any $\Deltak $ is also Hermitian. A matrix that is both Hermitian and symmetric has real entries, so every possible $\Deltak $ is a real and symmetric matrix.

This means that when we use the involution $\theta(g)=-g^T$ on the Lie algebra, all the Pauli strings in the Hamiltonian will be put into $\mathbf{m}$: $\theta(\Deltak )=-\Deltak ^T=-\Deltak $. So we can guarantee that a suitable Cartan decomposition of the lie algebra of the combinatorial Laplacian can always be found for any complex we choose to study.


\subsection{Details on the Cartan decomposition}
\label{app:cartan}
Our method relies on the simulation of the unitary evolution of the combinatorial Laplacian $\Deltak $. For this, we decide to make use of the Cartan decomposition for unitary operations initially proposed by Khaneja and Glaser~\cite{Khaneja_2001}, and we explicitly follow the algorithm of Kokcu et al.~\cite{Kokcu2022}. This method involves three main steps.

Firstly, having found the Pauli decomposition of the combinatorial Laplacian $\Deltak  = \sum_l c_l\hat{\mathcal{P}}_l$, we can find the closure under commutation of the Pauli terms, i.e. $\hat{\mathcal{P}}_l,[\hat{\mathcal{P}}_l,\hat{\mathcal{P}}_k], [\hat{\mathcal{P}}_l,[\hat{\mathcal{P}}_k,\hat{\mathcal{P}}_j]], \cdots$. This provides us with the basis for the Hamiltonian algebra, $\bm{\mathfrak{g}}(\Deltak )$, which is a subset of the full $\bm{\mathfrak{su}}(2^n)$ algebra.

The second step is to find a Cartan decomposition of the Lie algebra $\bm{\mathfrak{g}}$. This is defined as an orthogonal split $\bm{\mathfrak{g}}=\bm{\mathfrak{l}}\oplus\bm{\mathfrak{m}}$ which satisfy the commutation relations
\begin{equation}\label{cart_decomp}
    [\bm{\mathfrak{l}},\bm{\mathfrak{l}}]\subset \bm{\mathfrak{l}}, [\bm{\mathfrak{m}},\bm{\mathfrak{m}}] \subset \bm{\mathfrak{l}}, [\bm{\mathfrak{l}},\bm{\mathfrak{m}}] = \bm{\mathfrak{m}}
\end{equation}
This decomposition can be found using an involution. This is a homomorphism
\begin{equation}
    \theta: \bm{\mathfrak{g}} \rightarrow \bm{\mathfrak{g}} \quad \text{s.t.} \quad \theta(\theta(g))=g \quad \forall g \in \bm{\mathfrak{g}},
\end{equation}
and all commutators are preserved. It can be shown that by defining the subspaces by $\theta(\bm{\mathfrak{l}})=\bm{\mathfrak{l}}$ and $\theta(\bm{\mathfrak{m}})=-\bm{\mathfrak{m}}$, the commutation relations in Eq.~\eqref{cart_decomp} are automatically satisfied~\cite{Kokcu2022}.

In Ref.~\cite{D_Alessandro_2007}, it was shown that there is a one to one correspondence between Cartan involutions and Cartan symmetries (a symmetry $\Theta$ such that $\Theta^2=\mathds{1}$ up to a phase factor). An example Cartan symmetry is time reversal, which corresponds to the involution $\theta(g)=-g^T$. They also showed that for anti--unitary symmetries, such as time reversal symmetry, the corresponding involution places the symmetric elements of the Lie algebra into $\bm{\mathfrak{m}}$. As we demonstrate in the Appendix, the combinatorial Laplacian $\Deltak $ for any complex is always a real and symmetric matrix, hence it always has time reversal symmetry. Hence, the involution $\theta(g)=-g^T$ ensures that $\Deltak \in\bm{\mathfrak{m}}$. This is significant due to the KHK theorem. Given a Cartan decomposition $\bm{\mathfrak{g}}=\bm{\mathfrak{l}}\oplus \bm{\mathfrak{m}}$, define a Cartan sub--algebra $\bm{\mathfrak{h}}\subseteq \bm{\mathfrak{m}}$. Then for any $\hat{m}\in\bm{\mathfrak{m}}$, there exists a $\hat{K}\in e^{i\bm{\mathfrak{l}}}$ and a $\hat{h}\in\bm{\mathfrak{h}}$ such that
\begin{equation}
    \hat{m}=\hat{K}\hat{h}\hat{K}^\dagger .
\end{equation}
Thanks to the time reversal symmetry of $\Deltak $, we can always find a Cartan decomposition such that $\Deltak \in\bm{\mathfrak{m}}$. Hence, we can use the KHK theorem to write $\Deltak =\hat{K}\hat{h}\hat{K}^\dagger$, and the overall unitary evolution follows
\begin{equation}
    \hat{U}(t)=e^{-i\Deltak t} = \hat{K}e^{-i\hat{h}t}\hat{K}^\dagger
\end{equation}
Since our protocol requires finding the trace of this evolution operator, we can apply the cyclic properties of the trace to simplify our simulation, $\tr\{e^{-i\Deltak t}\} = \tr\{\hat{K}e^{-i\hat{h}t}\hat{K}^\dagger\}=\tr\{e^{-i\hat{h}t}\}$. We also note that $\mathfrak{h}$ is a Cartan sub--algebra, which is a maximal Abelian algebra in $\mathfrak{m}$. This means that every Pauli string in $\hat{h}\in\bm{\mathfrak{h}}$ commutes, so simulating $e^{-i\hat{h}t}$ is easy. Crucially, this means that the simulation error, and hence the circuit depth is independent of simulation time. This drastically reduces the number of gates required for time evolution compared to Trotterization.

The third and final step is to find the coefficients for each of the Pauli strings in $\hat{h}$. This is done via classical optimization: we minimize the function $f(\bm{\theta})=\langle\hat{K}(\bm{\theta})\hat{v}\hat{K}(\bm{\theta})^\dagger,\Deltak \rangle$, with $\hat{v}\in\bm{\mathfrak{h}}$ and $e^{it\hat{v}}$ dense in $e^{it\bm{\mathfrak{h}}}$. Once we find a local extremum of $f(\bm{\theta})$ denoted by $\hat{K}(\bm{\theta_c})$, we use the KHK theorem to find $\hat{h}$:
\begin{equation}
    \hat{K}(\bm{\theta_c})\Deltak \hat{K}(\bm{\theta_c})^\dagger = \hat{h} \in \bm{\mathfrak{h}}
\end{equation}
As discussed, the Cartan decomposition has the advantage in terms of quantum resources required for time evolution compared to Trotterization. However, calculating the coefficients for $\hat{h}$ can prove to be time consuming, as it requires many calculations of the function $f(\bm{\theta})$ and its gradient. Hence if minimizing the overall time for Betti number calculation is a higher priority than the minimization of quantum resources, simply performing multi--step Trotterization may be more ideal. The simulation error induced by finite--depth Trotterization would certainly hinder subsequent Betti number calculations, however.

Fig.~\ref{fig:lie_algebra}(a) shows the dimension of the Lie algebra, $|\bm{\mathfrak{g}}(\Deltak )|$, for all possible 4--vertex complexes. We observe a steady decrease in the Lie algebra dimension as we move from the near--complete complexes at low $k$, to the emptier complexes at high $k$. The consequence of this is shown in Fig.~\ref{fig:lie_algebra}(b); there is a clear correlation between the size of the Lie algebra, and the time taken to find the coefficients for $\hat{h}$. This was also explained in~\cite{Kokcu2022}, and this same behavior is found for larger complexes. Hence for a particular complex and homology group $k$, we can count the number of edges [relative to the maximum possible number, $\binom{n}{2}$], to give us a estimate of the order of magnitude of time required to calculate the Cartan parameters.

We know that the original $\Deltak $ lives in the computational basis reduced to $\Sk$. Subsequently, when calculating the trace of the evolution $\tr \{ e^{-i\Deltak t}\}$, we should only need to measure overlaps involving the computational basis states in $\Sk$, as shown in \eqref{eq:dos_overlap}. The states $\ket{\psi_j}\notin\Sk$ are orthogonal to the basis of $\Deltak$, so the overlaps; $\langle \psi_j| e^{-i \Deltak t} | \psi_j \rangle \approx \langle \psi_j| \mathds{1} | \psi_j \rangle -i\langle \psi_j| \Deltak | \psi_j \rangle + O(\Deltak^2)+\cdots = 1$. Therefore, the contribution of these states to the overall trace can be accounted for without any actual measurement. This applies with any approximate simulation of $\Deltak$ via Trotterization.

However, this is no longer true when the Cartan decomposition is applied; decomposing $\Deltak$ using the KHK theorem essentially induces a scrambling of the basis. Even though the reduced Hamiltonian $\hat{h}$ contains the same spectral information as $\Deltak$, the states $\ket{\psi_j}\notin\Sk$ now possess a non zero overlap with this new rotated basis of $\hat{h}$. Hence, Eq.~\eqref{eq:dos_overlap} does not apply when substituting $\Deltak$ for $\hat{h}$; and the measurements required for trace estimation in Sec.~\ref{sec:sampling} require sampling over all $2^n$ computational basis states, a sampling disadvantage particularly when $|\Sk|\ll2^n$.
\begin{figure}
    \centering
    \includegraphics[width=0.72\linewidth]{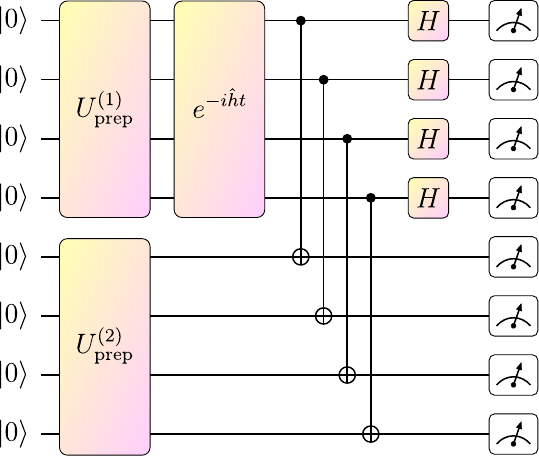}
    \caption{\textbf{Destructive SWAP test.} Example of destructive SWAP routine to obtain $|\langle 0|(\hat{U}_\mathrm{prep}^{(2)})^\dagger e^{-i\hat{h}t} \hat{U}_\mathrm{prep}^{(1)}|0\rangle|^2$ for $N=4$.}
    \label{fig:swapd}
\end{figure}

To demonstrate the benefits of the Cartan decomposition, we can take the simplicial complex in Fig.~\ref{fig:graphs}(a) as an example, with $k=0$. The Pauli decomposition of this complex's combinatorial Laplacian has 26 terms, but after doing the Cartan decomposition, we can find the reduced Hamiltonian $\hat{h}$ which possesses the same eigenspectrum as the full $\Deltak $. We find that $\hat{h}$ has just 8 Pauli terms, and the circuit required to simulate $e^{-i\hat{h}t}$ is given in Fig. \ref{fig:cart_circ}. This circuit has 35 gates, of which 16 CNOTs. Importantly, the number of gates is identical for all times $t$. On the other hand, if we simulated the evolution $e^{-i\Deltak t}$ using the first order Suzuki--Trotter expansion, just one Trotter step requires 96 gates and 40 CNOTs. Moreover, for larger simulation times and fixed accuracy, we would require an increasing number of Trotter steps to perform the DOS estimation, ultimately resulting in a gate count scaling with $t$.


\subsection{Destructive SWAP test}
\label{app:swapd}
The destructive SWAP test is a depth--2 circuit trading, with respect to the original SWAP test, an additional ancilla and CSWAP (Fredkin) gates for a linear scaling in $N$ of the post processing. In fact, every qubit of the destructive SWAP requires a final measurement.
We show an example destructive SWAP circuit in Fig.~\ref{fig:swapd}.
To evaluate $|\langle 0|(\hat{U}_\mathrm{prep}^{(2)})^\dagger e^{-i\hat{h}t} \hat{U}_\mathrm{prep}^{(1)}|0\rangle|^2$, we prepare the two states $\hat{U}_\mathrm{prep}^{(1)}|0\rangle$ and $(\hat{U}_\mathrm{prep}^{(2)})^\dagger|0\rangle$ on the two registers, perform $N$ number of CNOTs each of which connecting the first $N$ qubits $i$ with the second $N$ qubits $i+N$, and $N$ number of Hadamard gates on the control qubits. We finally read out both registers ($2N$ qubits). We call the two strings of measurement outcomes relative to the two registers $\bm{\psi}_{(1)}$ and $\bm{\psi}_{(2)}$. The destructive SWAP test fails when the bit--wise AND between the two strings of measurements has odd parity, i.e. $|\bm{\psi}_{(1)} \wedge \bm{\psi}_{(2)}|$ is odd. Therefore, we have $|\langle 0|(\hat{U}_\mathrm{prep}^{(2)})^\dagger e^{-i\hat{h}t} \hat{U}_\mathrm{prep}^{(1)}|0\rangle|^2 = 2(1-P_\mathrm{fail})-1$, where $P_\mathrm{fail}$ is the probability that the test fails.


\bibliographystyle{unsrtnat_compact} 
\bibliography{main}

\end{document}